# Multi-Band Mobility in Semiconducting Carbon Nanotubes

Yang Zhao,[1] Albert Liao,[1] and Eric Pop[1,2,*]

[1]*Dept. of Electrical and Computer Engineering, Micro and Nanotechnology Laboratory, University of Illinois, Urbana-Champaign, IL 61801, USA*

[2]*Beckman Institute, University of Illinois, Urbana-Champaign, IL 61801, USA*

We present new data and a compact mobility model for single-wall carbon nanotubes, with only two adjustable parameters, the elastic and inelastic collision mean free paths at 300 K. The mobility increases with diameter, decreases with temperature, and has a more complex dependence on charge density. The model and data suggest the room temperature mobility does not exceed 10,000 cm$^2$/V·s at high carrier density ($n > 0.5$ nm$^{-1}$) for typical single-wall nanotube diameters, due to the strong scattering effect of the second subband.

[*]Contact: epop@illinois.edu

## I. INTRODUCTION

Semiconducting single-wall carbon nanotubes (CNTs) are tightly rolled up sheets of graphene, with typical diameters 1–4 nm and band gaps 0.2–0.8 eV. As such, they present excellent electronic [1] and thermal properties [2], without the edge scattering associated with graphene nanoribbons of comparable band gap [3]. Experimental studies of CNT mobility have investigated diameter dependence [4], and probed its upper limits in long samples [5]. Previous theoretical efforts have developed sophisticated multi-band Boltzmann transport [6] and Monte Carlo [7] simulations of CNTs. However, while compact models have been introduced for ballistic conduction in CNTs [8], no such work exists for diffusive mobility calculations.

In this Letter we present a computationally inexpensive model for the low-field mobility in semiconducting CNTs, accounting for various phonon modes and multi-band transport. Analytic expressions are introduced for the scattering mean free paths (MFPs) with only two adjustable parameters, the elastic and inelastic MFP at 300 K. The mobility is calculated by integrating charge and averaging MFPs across multiple subbands, including effects of Pauli blocking. Good agreement is found with experimental data and previous Boltzmann transport simulations for mobility dependence on charge density, diameter and temperature. Interestingly, we find that several subbands are always required in transport at high charge densities ($n > 0.5$ nm$^{-1}$).

## II. MOBILITY MODEL

The one dimensional (1-D) field-effect mobility is $\mu = GL/(qn)$, where $G$ is the conductance, $L$ the device length, $q$ the elementary charge, and $n$ the carrier density per unit length. At low fields not far from equilibrium and in the relaxation time approximation, the 1-D conductance is [9, 10]:

$$G(T) = \frac{4q^2}{h} \sum_i \int_0^\infty \frac{\lambda(E,T)}{L + \lambda(E,T)} \left( -\frac{\partial f_0}{\partial E} \right) dE \qquad (1)$$

where $q$ is the elementary charge, $h = 2\pi\hbar$ the Planck constant, $\lambda$ the carrier MFP, $E$ the energy, $f_0$ the Fermi-Dirac distribution, and the summation index is over the participating subbands. This approach relates mobility with the (measurable) conductance, the carrier MFPs, and the carrier density. The latter is controlled capacitively through a gate voltage in experiments, setting the location of the Fermi level such that $n = \int g_S(E) f_0(E) dE$ where $g_S(E)$ is the semiconducting (S) nanotube density of states (DOS) [1]:

$$g_S(E) = g_M \sum_i \left(1 - \frac{\Delta_i}{E}\right)^{-1/2} u\left(E - \Delta_i\right).$$  (2)

Here $g_M = 4/(\pi \hbar v_F) \approx 2.08 \times 10^9$ eV$^{-1}$m$^{-1}$ is the single-band DOS of a metallic (M) nanotube, $v_F \approx 9.3 \times 10^5$ m/s is the Fermi velocity, and $u(E)$ the Heaviside unit step function. We include the first five subbands, with $\Delta_i = i(E_G/2)$ referenced to the middle of the gap, where $i = 1, 2, 4, 5, 7$ and plot the DOS in Fig. 1(a) [11]. The band gap itself is $E_G = (4/3)\hbar v_F/d \approx 0.82/d$ eV, where $d$ is the diameter in nanometers.

With this formalism in place, we now adapt the approach introduced in Ref. [12] for M-CNTs, to relate the microscopic scattering MFPs with the macroscopic mobility and conductance of S-CNTs. We recall that the optical phonon (OP) emission rate in M-CNTs at a reference temperature ($T = 300$ K) is proportional to the phonon occupation and DOS [12]:

$$\frac{1}{\tau_{M,OP-ems}(300)} = \frac{v_F}{\lambda_{OP,300}} \propto \left[N_{OP}(300) + 1\right] g_M$$  (3)

where $N_{OP} = 1/[\exp(\hbar\omega_{OP}/k_B T) - 1]$ is the OP occupation, $\hbar\omega_{OP} \approx 0.18$ eV and $k_B$ is the Boltzmann constant. The only independent parameter is now the OP emission MFP at 300 K, which was recently shown to scale with diameter as $\lambda_{OP,300} \approx 15d$ [13, 14]. In similar fashion, the OP scattering rate and MFP in S-CNTs can be written at any energy level or temperature as:

$$\frac{1}{\tau_{S,OP}(E,T)} = \frac{v_S(E)}{\lambda_{S,OP}} \propto \left[N_{OP}(T) + \frac{1}{2} \pm \frac{1}{2}\right] g_S(E \mp \hbar\omega_{OP})$$  (4)

where the upper (lower) signs correspond to emission (absorption), and the velocity $v_S = (1/\hbar)(\partial E/\partial k)$ is obtained consistently with the band structure in the DOS above. Combining Eqs. (3) and (4) yields an expression for the MFP of scattering with OPs in semiconducting nanotubes which takes into account their band structure, temperature, and carrier energy:

$$\lambda_{S,OP}(E,T) = \lambda_{OP,300} \frac{N_{OP}(300)+1}{N_{OP}(T)+\frac{1}{2}\pm\frac{1}{2}} \frac{g_M}{g_S(E \mp \hbar\omega_{OP})} \frac{v_S(E)}{v_F} \qquad (5)$$

where once again upper (lower) signs correspond to emission (absorption). A similar process is followed for acoustic phonon (AP) scattering, which is considered elastic since AP energies are typically $\ll k_B T$ [12]:

$$\lambda_{S,AP}(E,T) = \lambda_{AP,300} \frac{300}{T} \frac{g_M}{g_S(E)} \frac{v_S(E)}{v_F} . \qquad (6)$$

Here $\lambda_{AP,300} \approx 280d$ is the second independent parameter of our model, obtained by fitting against experimental data (Section III below). The combined MFP at any energy or temperature is obtained by Matthiessen's rule:

$$\frac{1}{\lambda} = \frac{1}{\lambda_{AP}} + \frac{1-f_0(E+\hbar\omega_{OP})}{\lambda_{OP,abs}} + \frac{1-f_0(E-\hbar\omega_{OP})}{\lambda_{OP,ems}}, \qquad (7)$$

where the subscript "S" was dropped. This is the MFP used to calculate the mobility in Eq. (1). Importantly, the Fermi occupation factors $(1-f_0)$ are included to account for Pauli blocking and ensure the final state after scattering is available [10].

## III. DISCUSSION

The computed MFPs are plotted against carrier energy in Fig. 1(b). Carriers can be either electrons or holes, due to band symmetry in CNTs, although hole mobility is experimentally easier to measure given better $p$-type contact metals [1, 4, 13]. Notably, the shape of the DOS has a profound effect on the MFPs, unlike in metallic nanotubes [12]. The elastic (AP) MFP "dips" correspond to DOS singularities at the beginning of each new subband. However, the sharp features of the OP absorption and emission MFPs occur $\pm\hbar\omega_{OP}$ above or below the locations of the singularities. The combined MFP (Eq. 7) is shown with a dashed black line, noting that OP emission ultimately plays only a small role in the low field mobility due to Pauli blocking. The derivative of the Fermi function ($\partial f_0/\partial E$ in Fig. 1(a) and Eq. 1) acts as a selection "window" for carrier energies that determine the mobility. The peak of this window is given by the Fermi energy location ($E_F$) as set by the gate-controlled charge density, and the width by the thermal broadening, $\sim 4k_B T$ [10].

In Fig. 2 we compare our model with the field-effect mobility extracted from conductance data of Ref. [4]. Mobility initially increases with carrier density, as the DOS in the first subband decreases, see Fig. 1(a). The mobility dips at higher charge density, when $E_F$ enters the second subband and opens up a strong new scattering channel. This is consistent with the experimental data and with previous simulations based on Monte Carlo and Boltzmann transport [6, 7], but it is successfully reproduced here in the context of an efficient, nearly-analytic model. As charge density increases further ($E_F$ advancing into the second subband), the DOS for scattering decreases once more (Fig. 1(a)), which results in a slight mobility rise at 54 K. We note this slight increase cannot be recovered when the second band is omitted.

Fig. 3(a) shows the charge density where $E_F$ enters the second subband for various diameter CNTs. Changing the CNT diameter modifies the band separations and subsequently the location of the mobility peak along with its magnitude, which must be kept in mind in practice. In Fig 3(b) we compare our model with the peak mobility data from [4], as well as experimental data we have similarly extracted in the course of this work at high charge density, $n \approx 1$ nm$^{-1}$ (a description of our CNT devices is provided in Ref. [13]). Lines represent model simulations for peak mobility, and at various charge densities as labeled. The *peak* mobility scales as a power law of CNT diameter, approximately $\sim d^2$, in part due to inverse scaling of the effective mass (band curvature) with diameter, and in part due to linear scaling of the MFPs with diameter [4, 13]. However, a more complex function of charge density emerges which cannot be captured on purely analytic grounds, but is easy to calculate in our framework. Our model and data suggest room temperature CNT mobility does not exceed 10,000 cm$^2$/V·s at high density ($n > 0.5$ nm$^{-1}$) for typical single-wall CNT diameters (1–4 nm), due to the strong scattering effect of the second subband.

Last but not least, it is relevant to comment on the effect of SiO$_2$ surface optical (SO) phonon scattering on CNT mobility, as suggested by recent theoretical work [15]. In principle, SO scattering should yield a stronger temperature dependence of mobility above 100 K. Within our model, SO phonons can be easily included similar to the inelastic OP scattering, with their own energy, occupation, and MFP fitting parameter at 300 K. However, we find our calculations reproduce existing CNT mobility data without inclusion of SO phonons. This does not necessarily imply that the SO phonon role is negligible, since existing mobility data may contain significant surface roughness or contamination to mask their effect (SO phonon scattering decays ex-ponentially with the CNT-substrate distance [15]). Clearly, more experimental work is needed to quantify the role of such scattering mechanisms in CNTs.

## IV. CONCLUSION AND ACKNOWLEDGEMENTS

We have presented an efficient mobility model for semiconducting CNTs with only two fitting parameters, the elastic and inelastic phonon scattering lengths at 300 K. Mobility dependence on diameter, temperature and charge density are reproduced in good agreement with experimental data and full Boltzmann transport simulations. This work is represents a practical approach which fills a gap of computational tools between analytic ballistic models [8] and full-band Monte Carlo simulations of diffusive transport [7].

We acknowledge valuable technical assistance from S. Dutta and discussions with Prof. F. Register. This work was supported in part by NSF-EMT and by the NRI MIND center.

**REFERENCES**

[1]     M. J. Biercuk, S. Illani, C. M. Marcus, and P. L. McEuen, "Electrical Transport in Single-Wall Nanotubes," in *Carbon Nanotubes, Topics in Applied Physics*, A. Jorio, G. Dressel-haus, and M. S. Dresselhaus, Eds.: Springer-Verlag, 2008, pp. 455-493.

[2]     T. Yamamoto, K. Watanabe, and E. R. Hernandez, "Mechanical Properties, Thermal Stability and Heat Transport in Carbon Nanotubes," in *Carbon Nanotubes, Topics in Applied Physics*, A. Jorio, G. Dresselhaus, and M. S. Dresselhaus, Eds.: Springer-Verlag, 2008, pp. 165-195.

[3]     X. Wang, Y. Ouyang, X. Li, H. Wang, J. Guo, and H. Dai, "Room-Temperature All-Semiconducting Sub-10-nm Graphene Nanoribbon Field-Effect Transistors," *Phys. Rev. Lett.,* vol. 100, 2008.

[4]     X. J. Zhou, J. Y. Park, S. M. Huang, J. Liu, and P. L. McEuen, "Band structure, phonon scattering, and the performance limit of single-walled carbon nanotube transistors," *Physical Review Letters,* vol. 95, p. 146805, 2005.

[5]     T. Durkop, S. A. Getty, E. Cobas, and M. S. Fuhrer, "Extraordinary mobility in semiconducting carbon nanotubes," *Nano Letters,* vol. 4, pp. 35-39, 2004.

[6]     V. Perebeinos, J. Tersoff, and P. Avouris, "Mobility in Semiconducting Carbon Nanotubes at Finite Carrier Density," *Nano Letters,* vol. 6, pp. 205-208, 2006.

[7]     G. Pennington, N. Goldsman, A. Akturk, and A. E. Wickenden, "Deformation potential carrier-phonon scattering in semiconducting carbon nanotube transistors," *Applied Physics Letters,* vol. 90, p. 062110, 2007.

[8]     D. Akinwande, J. Liang, S. Chong, Y. Nishi, and H.-S. P. Wong, "Analytical Ballistic Theory of Carbon Nanotube Transistors: Experimental Validation, Device Physics, Parameter Extraction, and Performance Projection," *J. Appl. Phys.,* vol. 104, 2008.

[9]     T. Fang, A. Konar, H. Xing, and D. Jena, "Mobility in semiconducting graphene nanoribbons: Phonon, impurity, and edge roughness scattering," *Phys. Rev. B,* vol. 78, p. 205403,

2008.

[10]    S. Datta, *Quantum Transport: Atom to Transistor*: Cambridge Univ. Press, 2006.

[11]    R. Saito, G. Dresselhaus, and M. S. Dresselhaus, "Trigonal warping effect of carbon na-
        notubes," *Physical Review B,* vol. 61, pp. 2981-2990, January 2000.

[12]    E. Pop, D. A. Mann, K. E. Goodson, and H. J. Dai, "Electrical and thermal transport in
        metallic single-wall carbon nanotubes on insulating substrates," *Journal of Applied Phys-
        ics,* vol. 101, p. 093710, May 2007.

[13]    A. Liao, Y. Zhao, and E. Pop, "Avalanche-Induced Current Enhancement in Semicon-
        ducting Carbon Nanotubes," *Physical Review Letters,* vol. 101, p. 256804, 2008.

[14]    V. Perebeinos and P. Avouris, "Impact excitation by hot carriers in carbon nanotubes,"
        *Physical Review B,* vol. 74, p. 121410R, September 2006.

[15]    V. Perebeinos, S. V. Rotkin, A. G. Petrov, and P. Avouris, "The Effects of Substrate
        Phonon Mode Scattering on Transport in Carbon Nanotubes," *Nano Letters,* vol. 9, pp.
        312-316, 2009.

[16]    S. Kar, A. Vijayaraghavan, C. Soldano, S. Talapatra, R. Vajtai, O. Nalamasu, and P. M.
        Ajayan, "Quantitative analysis of hysteresis in carbon nanotube field-effect devices," *Ap-
        plied Physics Letters,* vol. 89, p. 132118, September 2006.

**FIGURES**

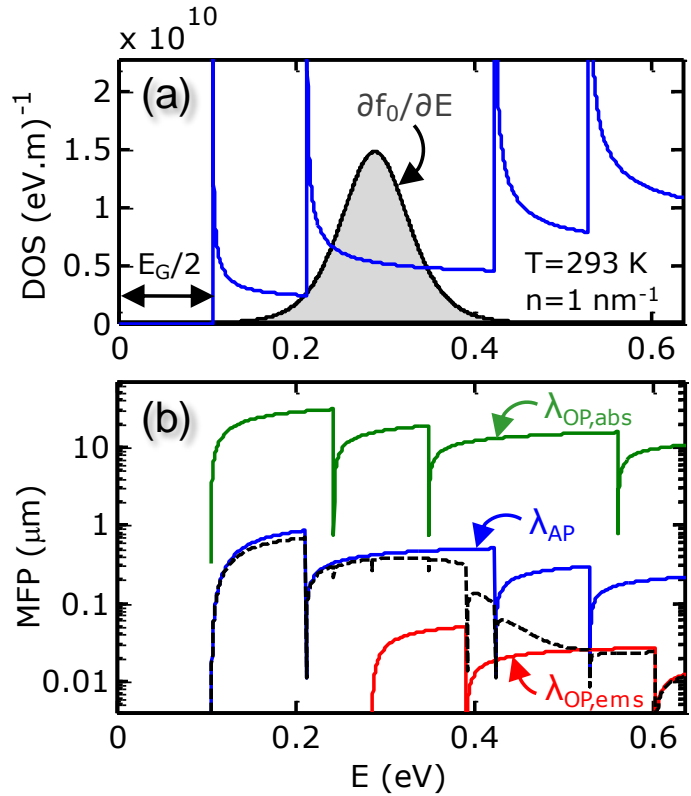

**Fig. 1:** (a) DOS of a CNT with diameter $d$ = 4 nm and $L$ = 4 μm, showing first four subbands, referenced to the middle of the band gap. Shaded area represents the derivative of the Fermi function. (b) Computed MFPs for the same CNT, with same energy axis. Dashed line is the combined MFP given by Eq. (7). Both figures at charge density $n$ = 1 nm$^{-1}$, highlighting the relevance of multiple subbands in CNT transport.

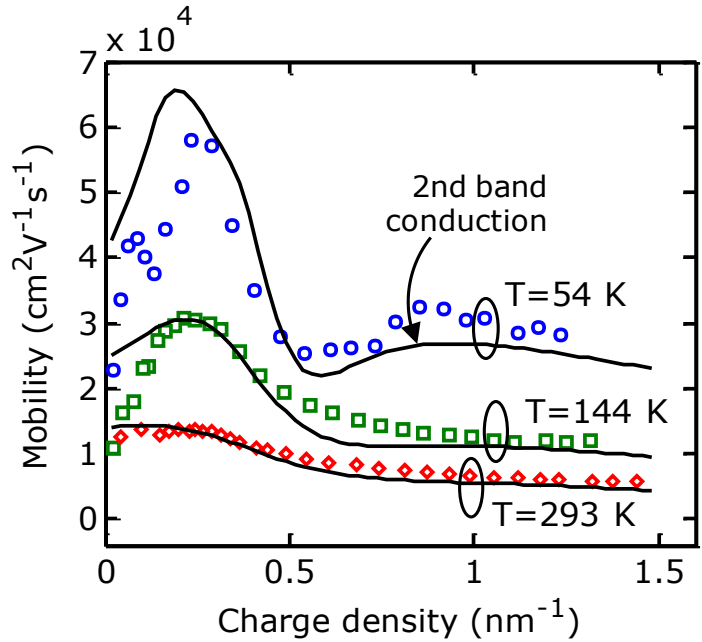

**Fig. 2:** Model (lines) and Ref. [4] data (symbols, $d \approx 4$ nm, $L \approx 4$ μm) at three temperatures. The Fermi level enters the second subband at 0.54 nm⁻¹ charge density, causing a sharp decrease in mobility due to the higher DOS for scattering. An upkick at low temperature (54 K) near 0.8 nm⁻¹ charge density is predicted and observable due to the slight charge contribution of the partially filled second subband.

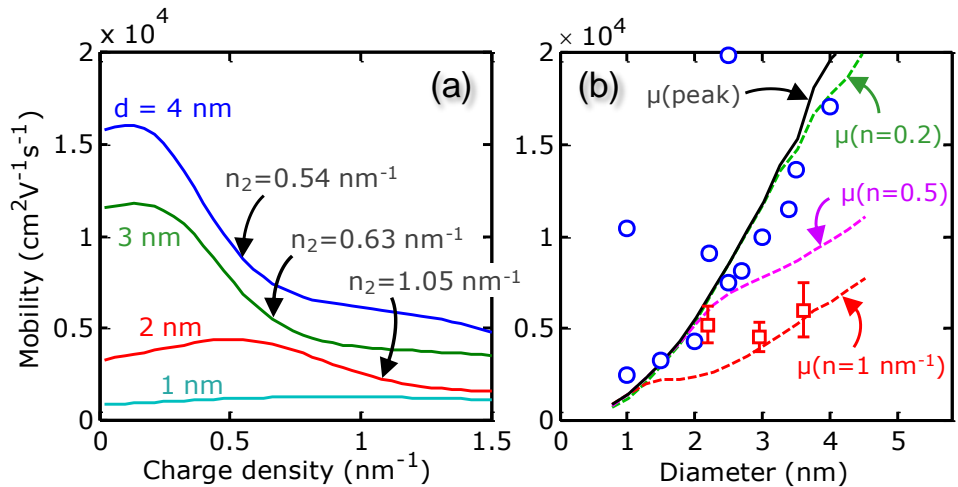

**Fig. 3:** (a) Computed mobility vs. charge density for various diameters ($T = 293$ K). Arrows indicate where Fermi level enters the second subband. (b) Computed mobility and experimental data vs. diameter. Peak mobility data from [4] (circles) includes their outliers. Our measurements (squares) at high charge density $n \approx 1$ nm$^{-1}$; error bars include uncertainty from forward/reverse hysteresis sweeps [16]. Black solid line is computed peak mobility, dashed lines are at charge densities as labeled.